\documentclass[final,5p,times,twoside,sort&compress]{elsarticle}
\usepackage{amsmath,amssymb}
\usepackage{color}
\usepackage{slashed}
\newcommand{\skipthis}[1]{}
 \def\Eq#1{Eq.~(\ref{#1})}
\def\Fig#1{Fig.~\ref{#1}}

\begin{document}
\begin{frontmatter}
\title{
{%
\vspace{-3.0cm}
\normalsize\hfill\parbox{32.0mm}{\raggedleft%
YITP-12-89
}}\\[0.5cm]
\vspace{2.cm}
Fluctuations in the quark-meson model for QCD with isospin
chemical potential} 

\author[label1]{Kazuhiko Kamikado}
\author[label2]{Nils Strodthoff}
\author[label2]{Lorenz von Smekal}
\author[label2,label3]{Jochen Wambach}

\address[label1]{Yukawa Institute for Theoretical Physics, Kyoto University, Kyoto 606-8502, Japan}
\address[label2]{Institut f\"{u}r Kernphysik, Technische Universit\"{a}t Darmstadt, 64289 Darmstadt, Germany}
\address[label3]{GSI Helmholtzzentrum f\"{u}r Schwerionenforschung GmbH, 64291 Darmstadt, Germany}

\begin{abstract}
\noindent 
We study the two-flavor quark-meson (QM) model with the functional
renormalization group (FRG) to describe the effects of collective mesonic
fluctuations on the phase diagram of QCD at finite baryon {\em and}
isospin chemical potentials,  $\mu_B$ and $\mu_I$. With only isospin
chemical potential there is a precise equivalence between the
competing dynamics of chiral versus pion condensation 
and that of collective mesonic and baryonic fluctuations in the
quark-meson-diquark model for two-color QCD at finite baryon chemical
potential. Here, finite  $\mu_B=3\mu$ introduces an additional
dimension to the phase diagram as compared to two-color 
QCD, however. At zero temperature, the  ($\mu_I,\mu $)-plane of this
phase diagram is strongly constrained by the ``Silver Blaze problem.''  
In particular, the onset of pion condensation must occur at  $\mu_{I}=
m_{\pi}/2$, independent of $\mu $ as long as $\mu + \mu_I$ stays below
the constituent quark mass of the QM model or the liquid-gas
transition line of nuclear matter in QCD. In order to maintain this
relation beyond mean field it is crucial to compute the pion mass
from its timelike correlator with the FRG in a consistent way. 
 \end{abstract}

\begin{keyword}
QCD phase diagram, isospin density, functional renormalization group. 
\end{keyword}
  \end{frontmatter}

\section{Introduction}

The phase diagram of Quantum Chromodynamics (QCD) continues to receive
enormous attention both experimentally and theoretically worldwide 
\cite{BraunMunzinger:2009zz,Friman:2011zz,Fukushima:2010bq}. QCD is
expected to reveal a rich phase structure in the plane of temperature $T$
and baryon chemical potential $\mu_{B}$. Monte-Carlo simulations 
in lattice gauge theory provide a powerful non-perturbative first
principle approach to QCD. At finite baryon chemical potential,
however, they are restricted by the fermion-sign problem, the fermion
determinant becomes complex, in general, and it can therefore no longer
be interpreted as part of a probability measure. There are various ways to
deal with the sign problem \cite{deForcrand:2010ys},
but the region of large baryon chemical potential and low temperature,
$\mu_B \gg T$, essentially remains inaccessible to lattice simulations. 
This is one motivation to study effective models with QCD symmetries 
such as (Polyakov-)Nambu-Jona-Lasinio ((P)NJL) 
\cite{Buballa:2003qv,Fukushima:2003fw,Ratti:2005jh} 
or (Polyakov-)quark-meson ((P)QM) models
\cite{Berges:1998sd,Braun:2003ii,Schaefer:2004en,Schaefer:2007pw},   
in order to describe the expected gross features of the 
QCD phase diagram at finite baryon chemical potential.

The sign problem also motivates studies of QCD-like theories without
this problem such as QCD with isospin chemical
potential $\mu_I$ or two-color QCD \cite{Son:2000xc,Kogut:2000ek}. In
both cases the fermion determinant remains real, and  
for an even number of degenerate quark flavors positive at finite
density. Consequently, lattice studies can be performed for both,  
QCD at finite isospin density
\cite{Kogut:2004zg,Sinclair:2006zm,deForcrand:2007uz,Detmold:2012wc} as well as 
two-color QCD at finite baryon density
\cite{Nakamura:1984uz,Hands:1999md,Hands:2000ei,Hands:2011ye}.  
In both these cases one studies a bosonic superfluid with
BEC-BCS crossover towards higher densities in the low temperature,
finite density region, however. An interesting QCD replacement with
fermionic baryons 
is the exceptional $G_2$ gauge theory \cite{Holland:2003jy}
which can also be simulated at finite density without sign
problem \cite{Maas:2012wr}. Such lattice simulations allow to
benchmark and cross-check the effective model calculations
\cite{vonSmekal:2012vx}  
as well as non-perturbative functional continuum methods 
for full QCD in general \cite{Braun:2009gm,Pawlowski:2010ht}.  

In this letter we use the functional renormalization group (FRG) to
study collective mesonic fluctuations in the two-flavor QM model for
QCD at finite baryon and isospin chemical potential. 
As compared to the standard ($\mu_B,T$) phase diagram, finite isospin
chemical potential thereby induces an imbalance between up and down
quarks as in neutron stars or heavy ion collisions, for example. 
The region of high baryon density, where this becomes relevant, cannot 
be described by a QM model without any baryonic degrees of freedom, of
course. On the other hand, in color superconducting quark matter
at even higher density, isospin chemical potential will eventually
destroy BCS pairing probably leading through an FFLO phase into an
unpaired state \cite{Fulde:1964zz,Alford:2000ze}. The other way round, 
finite baryon or isosymmetric quark chemical potential $\mu=\mu_B/3$
will lead to an imbalance of the Fermi spheres of up and anti-down
quarks and thus destroy pion condensation in the ($\mu_I,T$) phase diagram,
transgressing through an FFLO phase, likewise \cite{Son:2000xc}. 

Moreover, the standard QCD phase diagram with baryon chemical
potential is related to that with isospin chemical potential outside
the pion condensation phase via orbifold equivalence in the large
$N_c$-limit \cite{Hanada:2011ju}. Just as two-color QCD
\cite{Kogut:2000ek}, the latter can be described at sufficiently low
temperatures within the framework of chiral effective field theory
\cite{Son:2000xc,Son:2000by,Birse:2001sn} and random matrix theory
\cite{Klein:2003fy,Kanazawa:2011tt}. At zero temperature, chiral perturbation
theory predicts a second order quantum phase transition at 
$\mu^c_I = m_{\pi}/2$, the onset of pion condensation. This is an
exact property whose explicit verification from the grand partition
function is called the Silver Blaze problem \cite{Cohen:2003kd}.
It relates the onset of pion condensation to the (vacuum) pion mass
$m_\pi$. The charged pions couple to $2\mu_I$, so their excitation
thresholds are given by $m_\pi \pm 2\mu_I $, and the zero-temperature
limit of the partition function must remain independent of $\mu_I$
below the critical value of the isospin chemical potential at which  
$m_\pi - 2\mu_I = 0$. The two-flavor theory at $\mu_I = 0$  is
invariant under the full $O(3)$ flavor rotations in
$\pi_0,\pi_+,\pi_-$. Isospin chemical potential is associated with the
conserved charge corresponding to the $O(2)$ subgroup of rotations in
the $\pi_+,\pi_-$ plane.  It acts as an external field which
explicitly breaks the $O(3)$ down to $O(2)$, but the zero temperature
limit of the partition function remains unchanged as long as its
strength is not sufficient to excite charged pions. When the isospin
chemical potential reaches its critical value at  $\mu_I^c = m_{\pi}/2
$, on the other hand, the resulting pionic Nambu-Goldstone excitations
of zero energy lead to the spontaneous breaking of the corresponding
$O(2)$ symmetry, and Bose-Einstein condensation (BEC) of charged pions
occurs. 

Considering non-zero values for both isospin and isosymmetric quark
chemical potential, more general constraints arise from the Silver
Blaze property in the zero temperature ($\mu_I,\mu $) phase diagram as
we will discuss below. We furthermore present the phase diagram of the
QM model for two-flavor QCD in the three-dimensional parameter space 
of temperature, isospin and isosymmetric quark chemical potential. As
compared to previous model studies, mainly within the NJL model
\cite{Barducci:2004tt,Zhang:2006gu,Mukherjee:2006hq,Andersen:2007qv,Sasaki:2010jz,Mu:2010zz,He:2005nk,Xiong:2009zz,Ebert:2005cs},
the FRG approach is thereby  well equipped to include the
corresponding collective mesonic fluctuations in the description of
chiral versus pion condensation.   

\vspace{-.2cm}

\section{Quark-meson model with isospin chemical potential}

\vspace{-.1cm}

With the standard two-flavor chiral symmetry-breaking pattern,
$SU(2)_L \times SU(2)_R\rightarrow SU(2)_V$, finite isospin chemical
potential leads to a further explicit breaking of the $SU(2)_V$ isospin
symmetry down to the rotations about the iso-three direction, {\it
  i.e.}, on the quark level to $SU(2)_V \to U(1)_V^{(3)}$. 
In the pion condensation phase at $\mu_I \ge \mu_I^c(T) $, with
$\mu_I^c(0) = m_\pi/2$, this remnant chiral symmetry then
spontaneously breaks down to a discrete $Z_2$ symmetry,
$U(1)^{(3)}_V\rightarrow Z_2$, with a non-vanishing 
pion-condensate as the order parameter and one Nambu-Goldstone
boson. As the isospin chemical potential is further increased, the
pion condensate increases and the chiral condensate decreases
corresponding to a rotation of the vacuum alignment in a BEC-BCS crossover. 
The approximate chiral symmetry gets partially restored, and the
asymptotic $Z_2\times U(1)_A^{(3)}$ symmetry for $\mu_I\to\infty$ in
the pion condensation phase agrees with that obtained in a hypothetical
world with isospin chemical potential but without any chiral symmetry
breaking ($\chi$SB) and hence quark mass $m_q=0$ as summarized in
Table~\ref{tab:SSB}.

\begin{table}[ht]
 \begin{tabular}{cc||cc}
  & with\ $\chi$SB & without\ $\chi$SB \\
  \hline
  &$SU(2)_L \times SU(2)_R$& $SU(2)_L \times SU(2)_R$ \\
  $m_q \neq 0$ & $\downarrow$  &  $\downarrow$ & $\mu_I \neq 0$\\
  &$SU(2)_V$ & $U(1)^{(3)}_L\times U(1)_R^{(3)}$ \\
  $\mu_I \neq 0$&$\downarrow$ &  $\downarrow$&SSB\\
  &$U_{V}(1)$ & $Z_2\times U(1)_A^{(3)}$&    \\
  SSB&$\downarrow$  & \\
  &$Z_2$  & &\\
 \end{tabular}
 \caption{Symmetry breaking patterns in the charged pion condensation phase.}
 \label{tab:SSB}
\end{table}

In this letter we use the QM model as a chiral effective model in which
this symmetry pattern is realized. The Euclidean
Lagrangian of the QM model at finite quark chemical potential
$\mu=\mu_B/3$ but zero isospin chemical potential is given by 

\vspace{-1cm}
\begin{equation}
 \begin{split}
  {\cal L}_{\rm QM}&=  \bar{\psi} \left( \slashed{\partial} + g
    (\sigma + i \gamma_5  
  \vec{\pi}\vec{\tau})- \mu \gamma_0\right)  \psi \\
  &+\frac{1}{2}(\partial_{\mu} \sigma)^2  + \frac{1}{2}(\partial_{\mu}
 \vec{\pi})^2 + U(\sigma^2+\vec\pi^2)-c \sigma,
 \end{split}
\end{equation}
with mesonic potential $U(\sigma^2+\vec\pi^2)=\frac{m^2}{2} (\sigma^2
+ \vec{\pi}^2)+\frac{\lambda}{4!}(\sigma^2 
 + \vec{\pi}^2)^2$ and explicit $\chi$SB term $c\sigma$
 corresponding to a finite current quark mass. 
The Lagrangian is invariant under $O(3)$-isospin rotations. For the
iso-three direction these are of the form
\begin{equation}
\psi \rightarrow e^{i \tau_3\theta } \psi,\quad \psi^\dagger \rightarrow
 e^{-i \tau_3\theta } \psi^\dagger ,\quad\pi_\pm \equiv \pi_1 \pm i \pi_2
 \rightarrow e^{\mp i 2\theta} \pi_\pm , \;\;
\end{equation}
with an associated conserved current
\begin{equation}
\begin{split}
 J^3_{\mu}& = i\bar{\psi} \tau_3 \gamma_{\mu} \psi 
+ 2i (\pi_- \partial_{\mu}\pi_+ - \pi_+ \partial_{\mu}\pi_-).
\end{split}
\end{equation}
Isospin chemical potential is now introduced in the usual way, by
adding a term $\mu_I Q^3$ with the associated conserved charge to
the corresponding Hamiltonian. Following the standard derivation 
\cite{Kapusta:2006pm}, one obtains the Lagrangian with finite isospin
and isosymmetric quark chemical potential, conveniently written
in the basis of up and down quarks with $\psi=(\psi_u^T,\psi_d^T)^T$ as
\begin{equation}
\label{eq:Lagrangianqmmuf}
\begin{split}
 {\cal L}_{QM+\mu_I} =& \psi S_0^{-1}\bar \psi +\frac{1}{2}(\partial_{\mu}
 \sigma)^2  +\frac{1}{2}(\partial_{\mu}   \pi_0)^2 + U(\rho^2,d^2)- c \sigma \\ 
  &+ \frac{1}{2} \left((\partial_\mu + 2
    \mu_I\delta^0_\mu)\pi_+(\partial_\mu - 2\mu_I
    \delta^0_\mu)\pi_-\right),  
\end{split}
\end{equation}
where $\rho^2\equiv \sigma^2 + \pi_0^2$ (with $\pi_0\equiv \pi^3$),
$d^2 \equiv \pi_+ \pi_-$, and 
\begin{equation}
S_0^{-1}\!=\!\left(\begin{smallmatrix}
\slashed \partial+g(\sigma + i \gamma_5 \pi_0)-(\mu+\mu_I)\gamma_0 & g i \gamma_5 \pi_-\\
g i \gamma_5 \pi_+&\slashed \partial+g(\sigma - i \gamma_5 \pi_0)-(\mu-\mu_I) \gamma_0
\end{smallmatrix}\right)\!.
\end{equation}
Setting $\mu =0$ temporarily, it is evident in this representation
that there is a precise map between the quark-meson model for QCD with isospin
chemical potential $\mu_I$ and the corresponding 
quark-meson-diquark model for two-color QCD \cite{Strodthoff:2011tz}
with baryon chemical potential provided by the following
modifications: Instead of the bi-spinors in flavor here, one
introduces bi-spinors in color consisting of say a red quark $\psi_r$
and a green charge-conjugated antiquark $\psi^C_g \equiv \tau_2 C
\bar\psi^T_g$, with $\tau_2$ for complex conjugation in the flavor
$SU(2)$. The charged pions $\pi_-$ and $\pi_+$ are then replaced by a
scalar diquark-antidiquark pair, $\Delta$ and $\Delta^*$, coupled
to the baryon chemical potential $\mu_B = 2\mu$ for $N_c=2$ instead of
$2\mu_I$ for the charged pions in QCD with isospin chemical potential,
and the neutral pion $\pi_0$ is replaced by the isovector of three
pions $\vec \pi$, with $\pi_0 \to \vec\tau\vec\pi $ in $S_0^{-1}$, 
which increases the number of would-be-Goldstone bosons for
$\chi$SB from 3 to 5 in two-color QCD with two flavors. This
identification is valid only for the matter sector, of course, as the
gauge sectors of both theories are fundamentally different.

\section{FRG flow equations}

\subsection{Effective potential}

\begin{figure}[!ht]
  \includegraphics[width=0.97\columnwidth]{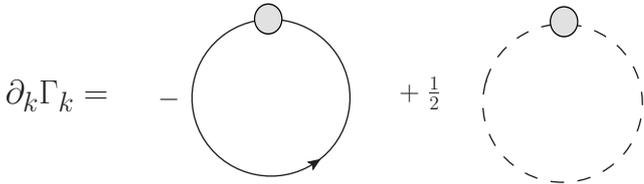}
  \caption{Diagrammatic representation of the flow equation for
    the effective action. 
    Solid (dashed) lines represent full field and RG scale $k$
    dependent fermion (boson) propagators and the circles insertions
    of $\partial_t  R_k(q)$.}
  \label{fig:picture_flow}
\end{figure}

The functional renormalization group is a well-established  
non-perturbative tool to study critical phenomena in quantum field
theory and statistical physics. It describes the evolution of the
scale-dependent effective average action $\Gamma_k$ from a microscopic
bare action specified at a UV cutoff scale $\Lambda_{UV}$ to the full
quantum effective action for $k\to 0$ in terms of a functional
differential equation \cite{Wetterich:1992yh} 
\begin{align}
\label{eq:floweq}
 \partial_k \Gamma_k[\phi]& =
 -{\rm Tr}\!\left[\frac{\partial_k R_{k \rm F}}{\Gamma^{(2,0)}_k+R_{k \rm F}} \right]
 +\frac{1}{2}{\rm
 Tr}\!\left[\frac{\partial_k R_{k \rm B}}{\Gamma^{(0,2)}_k+R_{k\rm B}}\right],
\end{align}
where $\Gamma_k^{(n,m)}$ denotes the $n\,\text{th}$ functional derivative
with respect to fermionic and $m\,\text{th}$ functional derivative with
respect to bosonic fields of the effective average action and the trace is
taken over internal degrees of freedom and momentum
space. \Fig{fig:picture_flow} shows a diagrammatic representation of
Eq.~(\ref{eq:floweq}). Although the flow equation has one loop
structure, it is exact.  Apart from the general requirements to act as
the desired infrared regulator consistent with the boundary conditions
\cite{Berges:2000ew,Pawlowski:2005xe}, 
there is a certain freedom in the precise choice of the 
fermionic/bosonic regulator functions $R_{k\rm F}$/$R_{k \rm B}$ which
can be exploited to minimize truncation effects. Here we use 
  \begin{align}
   R_{k \rm B}& = (k^2 - \vec{p}^2) \theta (k^2 - \vec{p}^2), \\
   R_{k \rm F}& =  \slashed{\vec{p}}\left(\sqrt{\frac{k^2}{\vec{p}^2}}-1\right) \theta
   (k^2 - \vec{p}^2) ,
   \label{eq:regulators}
  \end{align}
which are three-momentum analogues of the optimized regulator for the
local potential approximation (LPA) \cite{Litim:2001up}. They have the
particular advantage that the Matsubara sums in the flow can be
evaluated analytically. The regulators suppress the
propagation of momentum modes below the scale $k$. 
Quantum fluctuations from all momentum modes are included by integrating the
flow equation (\ref{eq:floweq}) from the bare classical
action at $\Lambda_{UV}$ down to $k=0$. From $n$ functional
derivatives of (\ref{eq:floweq}) one obtains the flow equations for
the $n$-point vertex functions which in turn contain up to $(n+2)$-point
functions. Just as Dyson-Schwinger equations, the flow equations for all 
$n$-point functions form an infinite hierarchy which requires
truncations to obtain a closed system of equations suitable for
non-perturbative solutions. 

One frequently used truncation scheme is the derivative expansion,
which has been applied successfully to $O(N)$ and quark-meson
models. In contrast to the NJL model, the QM model contains explicit
mesonic degrees of  freedom in the Lagrangian and is well suited 
to study mesonic fluctuations with the FRG. At leading order in the 
derivative expansion a scale-dependent effective potential $U_k$ is
obtained without wave function renormalizations and with
scale-independent Yukawa couplings. The Ansatz for the scale-dependent
effective action is chosen to comply with the expected 
symmetry-breaking pattern. In particular, because isospin chemical potential
explicitly breaks the $O(4)$ flavor symmetry of the mesonic sector 
down to $O(2) \times O(2)$, the effective potential $U_k$ must be
allowed to depend on two invariants,
$\rho^2=\sigma^2 + \pi_0^2$ and $d^2=\pi_1^2+\pi_2^2=\pi_+
\pi_-$. Therefore, our Ansatz for
the effective action is given by
\begin{equation}
\Gamma_k[\rho^2,d^2]=T\sum_n \int\text{d}^3 x \left.\mathcal{L}_{QM+\mu_I}\right|_{U\to U_k(\rho^2,d^2)},
  \label{eq:ansatzeffectiveaction}
\end{equation}
with $\mathcal{L}_{QM+\mu_I}$ from \Eq{eq:Lagrangianqmmuf}.
At the UV cutoff scale we assume an $O(4)$ symmetric potential of the form
\begin{equation}
U_{\Lambda_\mathrm{UV}}=a \left(\rho^2+d^2\right) +b \left(\rho^2+d^2\right)^2.
\end{equation}
Inserting the Ansatz (\ref{eq:ansatzeffectiveaction}) into the
Wetterich equation (\ref{eq:floweq}), the flow equation for the
effective potential can then be derived entirely analogously to that
of the quark-meson-diquark (QMD) model for two-color QCD
\cite{Strodthoff:2011tz}, yielding, 
\begin{equation}
 \begin{split}
 k \partial_k& U_k = \frac{k^5 }{12\pi^2} \Bigg[\frac{1}{E_{\pi}}
 \coth\left(\frac{{E_{\pi}}}{2T}\right)+\sum_{i=0}^2
 \frac{R_i}{\omega_i} \coth \left(\frac{\omega_i}{2 T} \right) 
  \\  &-\! N_c\!\!  \sum_{\pm} \frac{2}{E_\pm}\left(1\pm
  \frac{\mu_I}{E_q}\right) \left(\tanh
  \left(\frac{E_\pm+\mu}{2T}\right)\!+\!\tanh
  \left(\frac{E_\pm-\mu}{2T}\right) \right)\Bigg] .
 \end{split}
\label{eq:floweqeffectivepotential}  
\end{equation} 

\vspace{-.4cm}
\noindent
Here, $E_\pm$ denote $k$-dependent quark excitation energies, 
\begin{equation}
   E_{\pm}  = \sqrt{ g^2 d^2 + \big(E_q\pm\mu_I\big)^2} , \;\;\mbox{with}
   \;\; E_q=\sqrt{k^2+g^2 \rho^2}, 
\label{eq:1}
\end{equation}
and for the neutral pion, we introduced  
\begin{equation}
E_{\pi}=\sqrt{k^2+2 U_r},
\end{equation}
with $U_r\equiv \frac{\partial U_k}{\partial\rho^2}$,
$U_d\equiv\frac{\partial U_k}{\partial d^2}$, and  
$U_{rr}\equiv \frac{\partial^2 U_k}{(\partial\rho^2)^2}$, etc. 
used as shorthand notations for the derivatives of the scale-dependent
effective potential here and below. The $\omega_i$ are the positive
square-roots of the three poles in $\omega^2 = - p_0^2$ 
of the scale-dependent boson propagator in the $3\times 3 $ subspace
of $\pi^+$, $\pi^-$ and $\sigma $ mesons which mix at finite isospin
density. They are the roots of the same cubic polynomial as
given for the QMD model in \cite{Strodthoff:2011tz,vonSmekal:2012vx},
and the $R_i$ are the corresponding residues. For $d^2=\pi_+ \pi_- =
0$, without pion condensation and pionic fluctuations, one has
$\omega_0 = E_\sigma = \sqrt{k^2+2 U_r + 4 \rho^2 U_{rr}} $, $R_0 = 1$
for the sigma meson, and  $\omega_{1,2} = E_{\pi} \pm 2\mu_I $ with $R_{1,2} =
\omega_{1,2}/E_{\pi}$ for the charged pions.   
 
For zero isosymmetric quark or baryon chemical potential, $\mu=0$,
Eq.~(\ref{eq:floweqeffectivepotential}) is identical to
the corresponding flow equation for the effective potential of the QMD
model for two-color QCD  at finite baryon density
\cite{Strodthoff:2011tz} with $N_c=2$, $\mu_I \to \mu$, and a
multiplicity of 3 for the first term on the right in
Eq.~(\ref{eq:floweqeffectivepotential}) to account for the usual three
pions replacing the neutral one here.

\subsection{Pion pole mass and flow of the 2-point function}

It is common practice in quark-meson model studies to fix the
parameters by adjusting the eigenvalues of the Hessian of the
effective potential at its minimum to the physical meson masses.
We refer to these masses as the screening masses here.
They determine the spacelike static limit of the mesonic 2-point 
functions and thus the inverse of the corresponding
susceptibilities. These are not the physical masses, in general. 
With Bose-Einstein condensation of diquarks in two-color QCD or
charged pions here, however, the zero temperature onset at $m_\pi/2$
provides an exact definition of the physical pion mass which is manifestly 
different from its screening mass 
\cite{Strodthoff:2011tz}. The two differ 
by the amount by which the radiative corrections change when
extrapolating the pion propagator from $p^2=0$ to the pion pole at
$-p^2 = m_\pi^2$. 
This is all well-known, of course. Perhaps surprisingly, however,  
the independent exact mass definition via the
quantum phase transition shows that the standard mass
assignment in QM models can be off by as
much as 30\% \cite{Strodthoff:2011tz}. The
approximation to mistake the pion screening mass for the physical one
thus has a considerable Silver Blaze problem.

In general, for more realistic model parameters, independent of the
existence of a quantum phase transition, one therefore needs to
consider the poles in the zero-temperature propagators or, equivalently,
the zeros in the corresponding 2-point functions, to  
define the physical masses. They are determined from the asymptotic
behavior of the propagators at large Euclidean times. For the lowest
stable particle it suffices to solve analytically continued flow
equations at timelike total momenta, and find, {\it e.g.}, for the pion,     
\begin{equation}
\Gamma^{(0,2)}_{\pi}(p;\sigma_\mathrm{min})\bigg|_{- p^2 =
 m_{\pi}^2}  = 0.
 \label{eq:conditionpolemass}
 \end{equation}
At mean-field level, the Silver Blaze problem requires this 
pion pole mass to be computed from the random phase approximation
(RPA). It is then guaranteed to agree with the critical chemical
potential at the BEC quantum phase transition in NJL and QM models
\cite{Brauner:2009gu,Strodthoff:2011tz}. Here, we use the FRG to
calculate a pion pole mass in a way consistent with the derivative
expansion of the effective average action. Although an analogous flow
equation can be obtained for the 2-point function of the sigma, a pole
mass determination requires more work in that case and will not be
done here. The additional complication is associated with 
the finite width of the sigma meson from its decay into two pions,
which requires a more careful search for the sigma 
pole on the correct sheet in the complex plane.

\begin{figure}[!ht]
 \includegraphics[width=0.97\columnwidth]{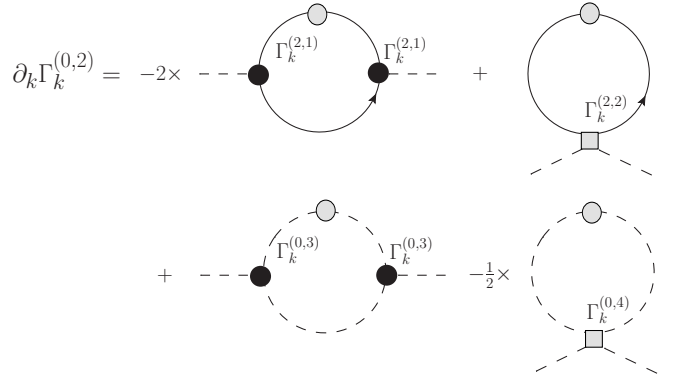}
 \caption{Diagrammatic representation of the flow equation for a mesonic
 2-point function.  
Solid (dashed)  lines represent scale-dependent fermion (boson) propagators and
  the light circles the insertions of $\partial_t R_k(q)$.}
 \label{fig:picture_2pt}
\vspace{-.2cm}
\end{figure}

Applying two functional derivatives to \Eq{eq:floweq}, one obtains 
the flow equations for the 2-point functions as
sketched in \Fig{fig:picture_2pt}, which depend on scale-dependent 3-
and 4-point functions with their own flow equations.   
Truncations are necessary in order to close this infinite tower of
equations for all $n$-point functions.

One example of such a truncation scheme is the so-called BMW
approximation \cite{Blaizot:2005wd, Blaizot:2006vr}. The basic idea is that,
because of the insertion of the regulator function, the dependence of
scale-dependent 3- and 4-point functions on the external momenta
is weaker than on the loop momentum. This motivates to expand 3- and
4-point functions in their external momenta. At leading order this
yields, {\it e.g.}, for the mesonic 3- and 4-point functions,
\[
 \Gamma^{(0,3)}_{ij l}(p,-p) =
 \frac{\partial\Gamma^{(0,2)}_{ij}(p;\phi)}{\partial \phi_l},\;\;
 \Gamma^{(0,4)}_{ij lm}(p,-p,0) = \frac{\partial^2
 \Gamma^{(0,2)}_{ij}(p;\phi)}{\partial\phi_m \partial \phi_l}. 
\]
The resulting closed coupled system of flow equations for the
2-point correlation functions has been studied for example for scalar
models in \cite{Guerra:2007vp,Benitez:2007mk}. Consistency
between the BMW and LPA approximations for vanishing external
momentum, where the 2-point function can also be obtained
from derivatives of the effective potential, is not guaranteed,
however. 

Therefore, we use an even simpler truncation which is manifestly
consistent with the leading order in the derivative
expansion for the effective average action. This is achieved
by using the scale dependent but momentum independent mesonic 
3- and 4-point vertices from the computation of the scale dependent
effective potential in the flow for the 2-point functions, {\it i.e.},
\begin{equation}
\Gamma^{(0,3)}_{k,\,ij l} \equiv
\frac{\partial^3 U_k}{\partial
  \phi_l \partial \phi_j \partial\phi_i}, \quad  
\Gamma^{(0,4)}_{k,\, ij lm} \equiv \frac{\partial^4 U_k }{\partial
  \phi_m \partial \phi_l \partial \phi_j \partial \phi_i}. 
  \label{eq:ansatz3and4ptfn}
\end{equation}
With constant Yukawa coupling at the leading order derivative
expansion the quark-meson vertices take their simple RPA forms as in
the NJL model \cite{Hatsuda:1985eb},    
\begin{align}
 \Gamma_{0}^{(2,1)} &= g ,\;\;\Gamma_{i}^{(2,1)} =  ig
 \gamma_5 \tau_i ,\;\;
 \Gamma_{ij}^{(2,2)}  = 0 .
\end{align}
The momentum dependence of the 2-point correlation function
now arises entirely from the external momentum through the
propagators in the loop. For the vacuum pion mass at zero isospin and
baryon chemical potential the effective potential remains $O(4)$ symmetric.
We therefore replace $U_k(\rho^2,d^2)$ by $U_k(\phi^2)$ with 
$\phi^2=\rho^2+d^2$ in our Ansatz for the effective average action.
The LPA-like flow of the pion 2-point function for Euclidean external 
momentum $p = (p_0 , \vec 0)$  then becomes,
\begin{align}
  k\partial_k \Gamma^{(0,2)}_{k,\, \pi}(p_0;\phi) =& \frac{k^5}{6 \pi^2} \Bigg(
-\frac{(N+1)U_k''}{E_{\pi}^3} -\frac{U_k'' +2 \phi^2
  U_k''' }{E^3_{\sigma}} + \notag\\
&\hskip -2.2cm  \frac{2U'' \big(E^2_{\sigma} -E^2_{\pi}\big)\left( (E_{\sigma}+E_{\pi})^3 (E^2_{\sigma}+E_{\sigma}E_{\pi}+E^2_{\pi})\!+\!(E_{\sigma}^3\!+\! E_{\pi}^3)p_0^2
 \right)}{E^3_{\pi} E^3_{\sigma}\big((E_{\pi}+E_{\sigma})^2 + p_0^2\big)^2}
\notag \\
& +\frac{8 N_f N_c g^2 (4 E_q^2- p_0^2)}{E_q\left(4 E_q^2+ p_0^2\right)^2}
\Bigg), \label{eq:floweq2ptrpapi} 
\end{align}
with 
standard primes for the $\phi^2$-derivatives here
in the place of the $\rho^2$ and $d^2$-derivatives of
$U_k(\rho^2,d^2)$ above, and $E_{\pi} =\sqrt{k^2 + 2U'}$, $E_{\sigma}
=\sqrt{k^2 + 2U' + 4 \phi^2  U''}$ correspondingly. We use $N=4$ for
the bosonic $O(4)$ symmetry with $N_f=2$ flavors (as compared to $N=6$
with $N_f=2$ in two-color QCD \cite{Strodthoff:2011tz}). At zero pion
momentum one easily verifies that \Eq{eq:floweq2ptrpapi} agrees with
the flow equation for $2U_k'$ obtained from
\Eq{eq:floweqeffectivepotential}.

We solve the flow equation for the real-time 2-point function by
analytically continuing $- p_0^2 = \omega^2 + i \varepsilon$ in
\Eq{eq:floweq2ptrpapi} before the integration of the flow
equation. The initial condition in the UV is then chosen to be 
\begin{equation}
\Gamma^{(0,2)}_{\Lambda_\mathrm{UV},\, \pi}(-i \omega ;\phi)=-\omega^2+2
U_{\Lambda_\mathrm{UV}}'(\phi^2) . 
\end{equation}

\subsection{Extended mean-field (eMF) approximation}

A worthwhile further simplification of the flow equations for
the effective potential in \Eq{eq:floweqeffectivepotential} and the
2-point function in \Eq{eq:floweq2ptrpapi} is to consider the purely
fermionic flow, {\it i.e.}, to neglect the mesonic contributions to
the flow. This is occasionally being referred to as {\em extended} mean-field
approximation (eMF) in the FRG literature because the fermionic flow
is independent of the mesonic potential. Moreover, it provides a
convenient way to include vacuum contributions which affect the usual 
mean-field studies \cite{Skokov:2010sf,Strodthoff:2011tz}. The
corresponding flow equations can then formally be integrated, yielding for the
effective potential at finite temperature, isospin and quark chemical  
potential, 
\begin{equation}
 \begin{split}
  U(\rho^2,d^2) &= U_\Lambda +   \frac{N_c}{6\pi^2}     
\sum_{\pm}\int_0^{\Lambda}\! d k \frac{k^4}{E_\pm}  \left(1\pm
  \frac{\mu_I}{E_q} \right) \, \times \\
  &\hskip 1.4cm  \left(\tanh\left( \frac{E_\pm+ \mu}{2
  T} \right)+\tanh\left( \frac{E_\pm-\mu}{2
  T} \right)\right),\\
  U_\Lambda &= a(\rho^2+d^2) + b\left(\rho^2+d^2\right)^2 -c \sigma -2
  \mu_I^2 d^2,  
\label{eq:effective action for MF}
\end{split}
\end{equation}
and for the vacuum pion 2-point function,
\begin{align}
  \Gamma^{(0,2)}_{\pi}(-i \omega;\sigma) &=  \frac{4g^2  N_c N_f}{3 \pi^2}\! \int_0^{\Lambda} \!\!dk k^4 
  \frac{4 E_q^2+\omega^2}{E_q\left(4
  E_q^2-\omega^2-i \varepsilon \right)^2}+ \Gamma^{(0,2)}_{\Lambda,\, \pi},\notag\\
  \Gamma^{(0,2)}_{\Lambda,\, \pi} &= -\omega^2+2 a+4 b \sigma^2.
  \label{eq:pion 2pt for MF}
 \end{align}
As before, the real-time result is obtained 
via analytic continuation with  $- p_0^2 = \omega^2 + i \varepsilon$
from the Euclidean correlation function.

\section{Results}

We solve the flow equations \Eq{eq:floweqeffectivepotential} and
\Eq{eq:floweq2ptrpapi} numerically on grids in field space.
For example, in the $O(4)$ symmetric case for $U_k(\phi^2)$ and
$\Gamma_{k,\,\pi}^{(0,2)}(p;\phi)$ this is a one-dimensional set of 
discrete field values of $\phi^2$. The necessary derivatives of the
effective potential can be obtained from finite differences or spline
interpolations. This leads to a system of ordinary differential
equations which can be integrated with standard methods. In comparison
to using Taylor expansions of the effective potential 
around a scale dependent minimum, it captures the form of
the effective potential everywhere on the grid in field space instead
of being restricted to some region around the minimum. 
It is thus well suited to describe first order phase transitions. Only 
recently this technique has been applied to problems with effective
potentials depending on several invariants
\cite{Strodthoff:2011tz}. We use the same techniques here when we
study the full two-dimensional flow from
\Eq{eq:floweqeffectivepotential} for $U_k(\rho^2,d^2)$ with collective 
fluctuations in both, the chiral and the charged pion condensate.


\subsection{Pole mass versus screening mass}

   \begin{figure}[t]
\leftline{\hskip .5cm \includegraphics[width=0.8\columnwidth]{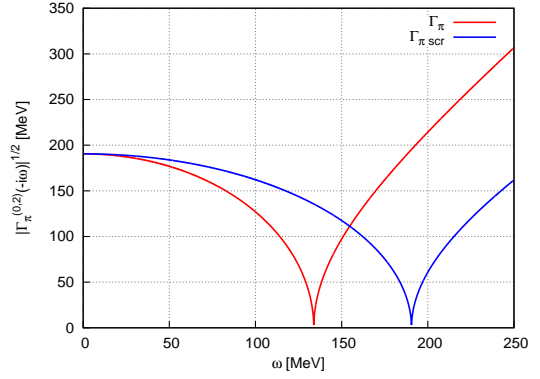}}
    \caption{Vacuum pion 2-point correlation function from the full
      FRG \Eq{eq:floweq2ptrpapi} as a function of timelike $\omega =
      ip_0$ compared to
    $\Gamma^{(0,2)}_{\pi,\mathrm{scr}}(-i\omega) \equiv -\omega^2 +
    {m_{\pi}^\mathrm{scr}}^2$ with the screening mass $m_{\pi}^{\rm
      scr}=\sqrt{2U'}$, both from the FRG parameter set C in
    Tab.~\ref{tab:parameter_2}.}  
    \label{fig:correlation}
   \end{figure}

 \Fig{fig:correlation} shows the square root of the modulus of the 
pion correlation function evaluated at the minimum of the effective
potential $\sigma_\mathrm{min}$ as a function of timelike 
external $\omega = ip_0$. For comparison, the corresponding 
 $\Gamma^{(0,2)}_{\pi,\mathrm{scr}} \equiv -\omega^2 +
 {m_\pi^\mathrm{scr}}^2$ with the screening mass $m_\pi^\mathrm{scr}$
 from the curvature of effective potential is also shown. 
 Pole and screening mass are the zeros of
 $\Gamma^{(0,2)}_{\pi}(-i\omega;\sigma_\mathrm{min})$ and
 $\Gamma^{(0,2)}_{\pi,\mathrm{scr}}(-i\omega;\sigma_\mathrm{min})$. Moreover,
 from Eqs.~(\ref{eq:floweqeffectivepotential}) and
 (\ref{eq:floweq2ptrpapi}),  ${m_\pi^\mathrm{scr}}^2  = 
 \Gamma^{(0,2)}_{\pi,\mathrm{scr}}(0;\sigma_\mathrm{min}) =
 \Gamma^{(0,2)}_\pi(0;\sigma_\mathrm{min}) $, in our truncation.  
 
In the present example, the pole mass is $m_\pi^\mathrm{pole}
= 133.0$~MeV which is within 3\% of the correct value as defined 
by the quantum phase transition at $2\mu_I^c=136.6$~MeV with the
parameters used here. The screening mass, on the other hand, 
is much heavier, with $m_\pi^\mathrm{scr}= 188.0$~MeV it
overestimates the correct value by about 38\%. The Silver Blaze
relation $m_\pi = 2\mu_I^c$ is satisfied exactly  for the 
pion pole mass in the eMF calculation, as it must
\cite{Strodthoff:2011tz}. For a full FRG 
calculation, the pion pole mass in our truncation is close and 
represents a considerable improvement as compared to the screening
mass. Perfect agreement should not be expected in this calculation for
the pion pole mass either, since the truncation for the flow of the
2-point function goes beyond the leading order derivative
expansion. One possible further improvement, for example, 
would therefore be to feed the momentum
dependent propagators back into the flow equation for the effective
potential and obtain both in an iterative procedure.   
Maybe more importantly, however, these results once more 
demonstrate that assigning physical values to the screening masses is 
a rather poor way of fixing model parameters. This is even more so for
the pion mass with 3 colors here than it is in the
corresponding 2-color calculation \cite{Strodthoff:2011tz}.

Less pronounced, the same inconsistency between screening mass and
quantum phase transition at $\mu=\mu_I^c$ has been observed
already in a purely bosonic model with isospin chemical potential
\cite{Svanes:2010we}. For a screening mass of
$m_\pi^\mathrm{scr}=138$~MeV, the zero-temperature onset of pion
condensation is observed in this model at $2\mu_I^c=132.6$~MeV,
corresponding to a moderate mismatch of 4\%. For comparison, we obtain a pion
pole mass of $m_\pi^\mathrm{pole}=129.9$~MeV in our truncation for
this purely bosonic model. The resulting deviation of only 2\% again 
represents at least some improvement as compared to the screening
mass. The relatively small deviation of pole and screening masses here
is in line with the general observation that fermions provide the
dominant contributions to the flow in QM models.

\subsection{Parameter fixing}

\paragraph{eMF calculation} For the extended mean-field calculation, we select
parameter sets A, B, C and D which reproduce the pion decay constant
$f_{\pi}$ and the pion pole mass and in the vacuum from
Eqs.~(\ref{eq:effective action for MF}) and (\ref{eq:pion 2pt for
  MF}), as summarized in Tab.~\ref{tab:parameter 1}. In this range of
parameters the mass $m_{\sigma}$ of the sigma meson increases
monotonically with the quartic coupling $b$ in the UV potential.

\begin{table}[ht]
\begin{center}
 \begin{tabular}{|c||c|c|c||c|c|c|}
  \hline
   & $a/{\Lambda^2}$ & $b$ & ${c}/{\Lambda^3}$&$f_{\pi}$ &$m^\mathrm{pole}_{\pi}$&$m^\mathrm{scr}_{\sigma}$\\
  \hline
  A &0.245 &1.0 &0.0124 &92.8 &138 &457\\
  \hline
  B &0.196 & 2.0&0.0123 &92.9 &137 &504\\
  \hline
  C & 0.09&4.0 &0.0125 &92.4 &138 &698\\
  \hline
  D & -0.09&8.0 &0.0125 &92.8 &139 &868\\
  \hline
 \end{tabular}
\end{center}
\vspace{-.4cm}
\caption{Parameters for eMF
 calculations with UV cutoff $\Lambda=600$~MeV and Yukawa coupling
 $g=3.2$; $f_\pi$ and meson masses are given in MeV.} 
 \label{tab:parameter 1}
\vspace{-.2cm}
\end{table}

\begin{figure}[t]
\leftline{\includegraphics[width=0.5\columnwidth]{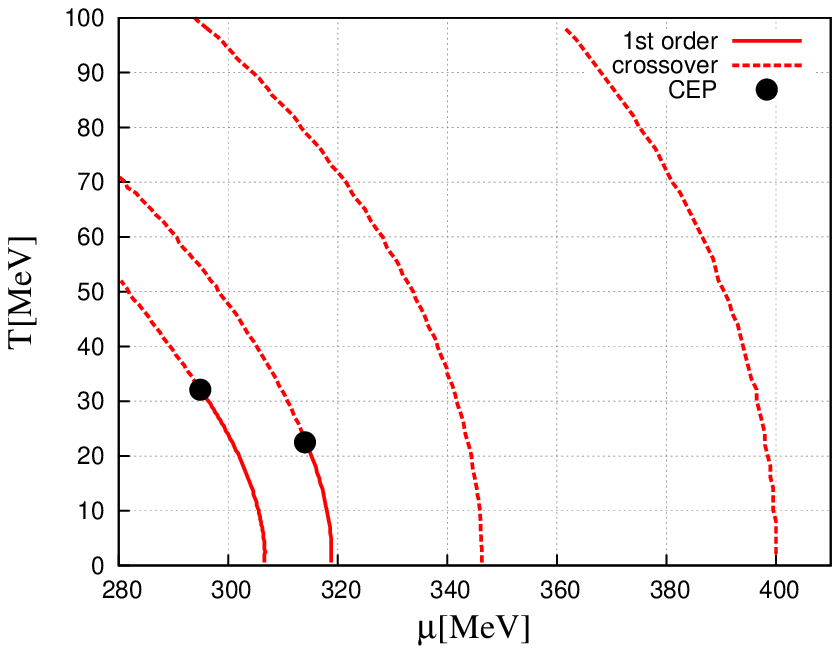}
  \includegraphics[width=0.5\columnwidth]{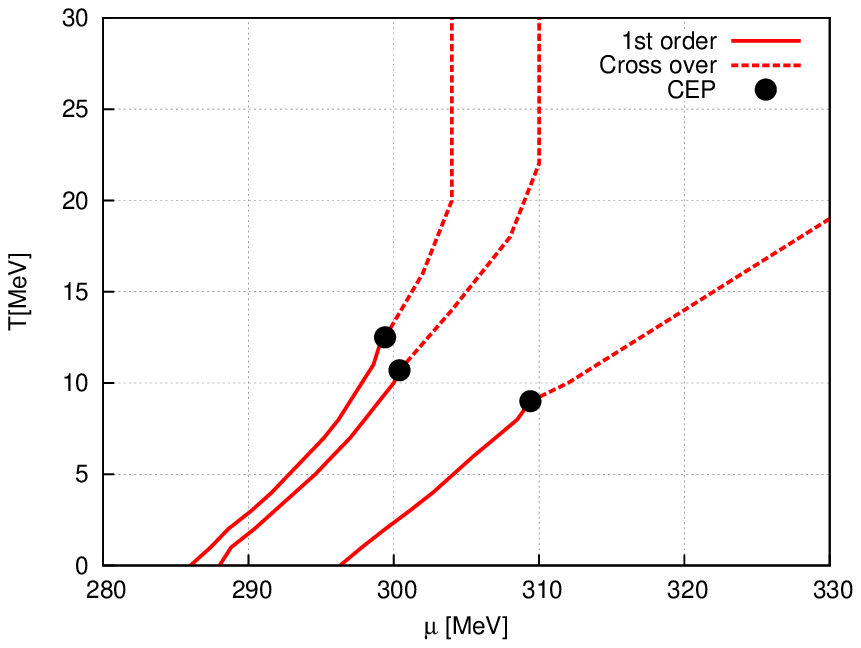}}
  \caption{QM model phase diagram at $\mu_I=0$ from eMF (left,
    parameter sets A,B,C, D from left to right) and full FRG
    (right, parameter sets A,B,C) calculations. Solid lines represent
    first order phase boundaries and dashed lines trace minima in the
    screening mass of the sigma meson to indicate chiral crossovers.}
  \label{fig:phase_diagram_muf=0}
\end{figure}

The left panel in \Fig{fig:phase_diagram_muf=0} shows the resulting 
($\mu,T$) phase diagram with $\mu_I = 0$ and no pion
condensation. For the parameter sets A and B there is a critical
endpoint (CEP) at $(T,\mu)=$ (22.5~MeV, 314~MeV) and (32~MeV, 295~MeV),
respectively. The position of the CEP is very sensitive to the choice of
parameters. With increasing $m_{\sigma}$ the CEP moves to lower
temperatures and eventually disappears from the phase diagram beyond a critical
value of $m^\mathrm{scr\, *}_{\sigma} \sim 550$~MeV, so there is none for
parameter sets C and D. A similar behavior was observed in a
mean-field 3-flavor QM model calculation  \cite{Schaefer:2008hk}. An
intuitive understanding of this effect is gained by considering  
a Landau expansion of the effective potential in  $\sigma^2$, yielding,
\begin{equation}
 \begin{split}
  U(\sigma,T,\mu)& = a'(T,\mu)\sigma^2 +  b'(T,\mu)\sigma^4 +  d(T,\mu) \sigma^6, \\
  a'(T,\mu) &= a(T,\mu) + a,\quad b'(T,\mu) = b(T,\mu) + b .
\label{eq:Landau_expansion}
 \end{split}
\end{equation}
To describe a first order phase transition we need to keep terms up to sixth 
order. The corresponding quark contributions are $a(T,\mu)$,
$b(T,\mu)$ and $d(T,\mu)$. They are independent of the initial
parameter choices for $a$ and $b$. Around the first order phase 
boundary, the total coefficients satisfy $a' > 0$, $b' < 0$ and $d > 0$,
which is just the condition for a double-well. For a sufficiently
large quartic bare coupling $b$, this condition can no longer be
satisfied. Thus the first order phase transition and the CEP have to
disappear from the ($\mu,T$) phase diagram as $m_{\sigma}$ increases.

The UV cutoff scale of $\Lambda=600$~MeV, which we use for the eMF
calculation here, allows a most direct comparison with earlier
mean-field results. From these it is known that vacuum contributions have 
a considerable effect on the chiral first order line and CEP
\cite{Skokov:2010sf}. For standard mean-field calculations vacuum
terms with a sharp 600 MeV cutoff were found to well reproduce
cutoff-independent results in dimensional regularisation 
\cite{Strodthoff:2011tz}. The eMF results include such
contributions by construction. They are therefore best compared to those
earlier calculations when one starts the fermionic flow at 600 MeV also.     

However, while eMF results for a given fixed sigma mass are
practically cutoff independent, sigma masses as low as
those with sets A and B in Tab.~\ref{tab:parameter 1} are not possible 
for larger UV cutoff scales such as the 900 MeV used in the FRG
results below. This is due to an emerging instability in the UV
potential, with $b<0$ in (\ref{eq:Landau_expansion}). As a result, we
are then unable to reach sigma masses below the $m^\mathrm{scr\,
  *}_{\sigma} \sim 550$~MeV needed for a chiral first order transition
and CEP. For a comparison with the full FRG results below it
therefore also seems more appropriate to keep the lower cutoff in the eMF
calculations. The main focus of this letter, the pion condensation
phase at finite isospin chemical potential, however, is not affected
by this choice. The eMF results in this region of the phase diagram
remain essentially unchanged when we go to the 900 MeV UV cutoff scale. 

\paragraph{FRG calculation}
Three sets for initial parameters in full FRG solutions are listed in  
Tab.~\ref{tab:parameter_2}. As for the eMF results,
the screening mass of the sigma monotonically increases with the
quartic coupling parameter $b$.

\begin{table}[!ht]
\begin{center}
 \begin{tabular}{|c||c|c|c||c|c|c|}
  \hline
   & $a/{\Lambda^2}$ & $b$ & ${c}/{\Lambda^3}$&$f_{\pi}$ &$m^\mathrm{pole}_{\pi}$&$m^\mathrm{scr}_{\sigma}$\\
  \hline
  A & 0.3224&0.25 &0.0045 &93.3 &134.3 &524.1\\
  \hline
  B &0.2844 & 1.25&0.0045 &92.8 &134.1 &557.1\\
  \hline
C&0 &9.075 &0.0045 &92.8 &133 &696.7\\
  \hline
  \hline
  A'&-0.48&9.74&0.0024&93.0&138.0&448\\
  \hline
 \end{tabular}
\end{center}
\vspace{-.4cm}
\caption{Initial parameters for full FRG calculations (parameter set
  A' for the purely bosonic model). The UV cutoff 
 is set to $\Lambda=900$~MeV and $g=3.2$; $f_{\pi}$ and meson masses
 are given in MeV.} 
\label{tab:parameter_2}
\end{table}


The right panel in \Fig{fig:phase_diagram_muf=0} shows the $\mu_I=0$ phase
diagram in the ($\mu,T$) plane in the vicinity of the CEP 
for the three parameter sets A,B and C. In contrast to the eMF
calculation, the phase structure for all sets is  
quite similar, although the screening masses are rather different.
In all cases there are first order phase transitions and CEPs.
In the full FRG effective potential, it makes no sense to separate the
initial potential from the contributions due to fluctuations as in
\Eq{eq:Landau_expansion}. Differences at the UV scale are reduced by
the mesonic fluctuations. As a result, the phase structure from full
FRG calculations is less sensitive to the resulting sigma mass than
in mean-field calculations.

\subsection{Zero temperature phase diagram}

\begin{figure}[t]
\vspace*{-.2cm}
  \includegraphics[width=0.97\columnwidth]{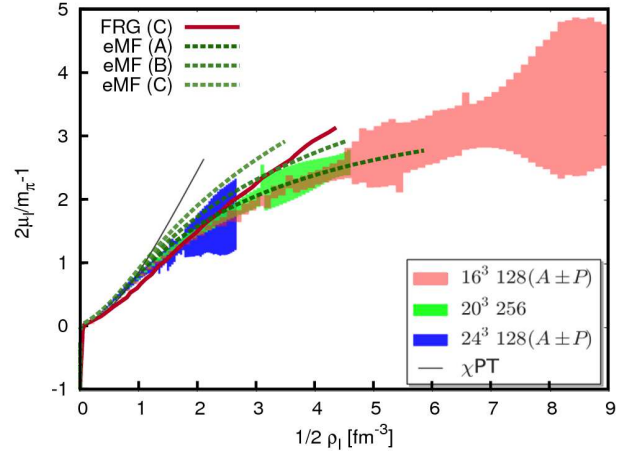}
\vspace{-.4cm}
  \caption{Isospin chemical potential over isospin density at 
    zero temperature and baryon chemical potential in comparison to
    the lattice data from Ref.~\cite{Detmold:2012wc}. The isospin
    densities are rescaled here to adjust to the
    lattice parameters. This mainly compensates for 
    the larger pion mass and decay constant on the lattice.}      
  \label{fig:density}
\end{figure}

First consider the zero temperature quantum phase transition with pion
condensation at $\mu= 0$. We have calculated the isospin density from
the $\mu_I$-derivative of the effective potential. The result is shown
in \Fig{fig:density} where the rescaled isospin chemical
potential $2\mu_I/m_\pi $ is plotted over the isospin density $\rho_I$
for our eMF and FRG solutions in comparison with very recent
lattice data from Ref.~\cite{Detmold:2012wc} and the leading order
chiral effective field theory ($\chi$PT) result \cite{Son:2000xc}.
The isopin density remains zero below the onset of pion condensation at
$\mu_{I}^c=m_\pi/2$ in the eMF calculation as it must while the full FRG result 
shows a slight residual Silver Blaze problem here, which might be due
to numerical integration and infrared cutoff
uncertainties. 
For large $\mu_I$, deep into the pion condensation phase, it gets
increasingly difficult to control systematic errors in the full FRG
calculations, {\it e.g.}, because the assumption of $\mu_I$-independent
initial parameters at a fixed ultraviolet cutoff breaks down eventually.
 
While the chiral effective field theory result describes the BEC phase
well, it appears to miss some essential dynamics in the BEC-BCS
crossover, which we estimate to occur around $2\mu_I/m_\pi - 1 \approx
2/3 $ in our calculations by the simple criterion that the quarks'
Dirac mass falls below $\mu_I$ there. Because the
vacuum realignment in this crossover is characterized by the mixing
between the sigma meson with the charged pions in the QM model, this
suggests that it might indeed be this mixing of the meson mass
eigenstates which is responsible for the observed inflection point and
back-bending of the isospin density at large $\mu_I$. The analogous
mixing is seen in the mass spectrum of the QMD model for two-color QCD
\cite{Strodthoff:2011tz}, between diquarks and the sigma
meson. Basically, the latter is infinitely heavy in $\chi$PT, so the
low-lying excitation spectrum of the QM model is essentially different
from that of the non-linear sigma model, especially in the crossover
region where the mixing occurs. The qualitative effect on the isospin
density can be illustrated in the {\em linear} sigma model, where a simple
calculation yields, 
\begin{equation}
\rho_I(x,y) =2f_\pi^2 m_\pi \,x\, \left(\frac{y^2-3}{y^2-1}-\frac{1}{x^4}+\frac{2}{y^2-1}x^2\right),
\end{equation} 
for $x\equiv 2\mu_I/m_\pi\geq 1$, and with $y\equiv m_\sigma/m_\pi$.
In the  $y\to\infty$ limit, this reduces to the $\chi$PT result $ \rho_I(x)
=2f_\pi^2 m_\pi \,x\,\big(1-x^{-4}\big)$  \cite{Son:2000xc}. Compared to
that, a finite sigma mass $y<\infty$ in the linear sigma model 
has the same qualitative effect as observed in the eMF results of
\Fig{fig:density} for parameter sets A, B and
C in Tab.~\ref{tab:parameter 1}.

The general structure of the phase diagram in the $T=0$
plane is also strongly constrained by the Silver Blaze property. 
The key to its understanding is the fact that different degrees of
freedom couple to different combinations of the chemical
potentials. While up(down)-quarks couple to $\mu\pm\mu_I$, charged
pions obviously only couple to the isospin chemical potential
 $\mu_I$. The partition function and, correspondingly, thermodynamic
 observables must remain independent of $\mu$ {\em and} $\mu_I$ as
 long as $\mu+\mu_I<m_q \equiv g\sigma_\mathrm{min} = gf_\pi $ 
and $\mu_I<m_\pi/2$, where $m_{q}$ and $m_\pi$ are the
vacuum quark and pion masses, respectively. This defines a
quadrilateral area in the zero-temperature phase diagram, which we
refer to as the {\em first Silver Blaze region} in the
following. The physical picture is that in absence of bound quark
matter, the horizontal line in the ($\mu_I,\mu$) plane defines the  
boundary of this region beyond which a degenerate Fermi gas of
up-quarks forms, just as the vertical line marks the onset of pion
condensation.  Another constraint arises inside the pion
condensation phase: For $\mu_I> m_{\pi}/{2}$ the vacuum changes and
the grand potential becomes $\mu_I$-dependent, of course.
Nevertheless, for constant $\mu_I$, it has to remain independent
of the isosymmetric quark chemical potential 
$\mu $ as long as this stays below the lightest quark
mass, {\it i.e.}, for $\mu < m^-_q(\mu_I)$, where 
\begin{equation}
m_q^\pm (\mu_I)=\sqrt{g^2 d^2+(g\rho\pm \mu_I)^2}
\end{equation}
are the quark masses at $\mu=0$. They no longer correspond to
pure up- and down- quark excitations as these mix in the pion 
condensation phase, and the quark masses are obtained by
diagonalizing $\Gamma^{(2,0)}_k$. More generally, from the quark
dispersion relation 
\begin{equation} 
E_q^-(\vec p^2)=\sqrt{g^2 d^2+\left(\sqrt{\vec p^2+g^2\rho^2}- \mu_I\right)^2},
\end{equation}
one furthermore notes that for $\mu_I > g\rho$ it becomes energetically
favorable to excite quarks with a finite spatial momentum
 $\vec p^2=\mu_I^2-g^2\rho^2$ and minimal energy
 $E^-_\mathrm{min}=g d$ \cite{Ebert:2005cs}. This might be an indication for
an inhomogeneous phase at large isospin chemical potential such as
the FFLO phases \cite{Fulde:1964zz,Alford:2000ze} also discussed for
two-color QCD with isospin chemical potential
\cite{Fukushima:2007bj,Andersen:2010vu} or in the 1+1 
dimensional NJL model \cite{Ebert:2011rg}. 

Therefore, the conditions $\mu_I>m_\pi/2$ with $\mu < m^-_q(\mu_I)$
for $\mu_I<g\rho $, or $\mu < E^-_\mathrm{min}$ for $\mu_I>g\rho$, together
 define a region in the zero-temperature phase diagram in which
the $\mu_I$-dependent grand potential still remains independent of $\mu$,
nevertheless. We will call this the {\em second Silver Blaze region}
in the following. 

Note however that the whole argument holds only as
long as there are no first order transitions in these regions of the
phase diagram. This concerns in particular the possible first 
order transition for large $\mu$ and small $\mu_I$ which may intersect
with the first Silver Blaze region. If it does, the CEP where it ends
must lie outside, however. This could then be the chiral first order
transition or a liquid-gas transition to bound quark matter below the
threshold for free quarks, for example, which are of course not
excluded. In contrast, a first order line with an endpoint inside the
Silver Blaze region would be thermodynamically
inconsistent. It is reassuring that we never observe this in  
our numerical calculations either.

 \begin{figure}[t]
  \includegraphics[width=0.97\columnwidth]{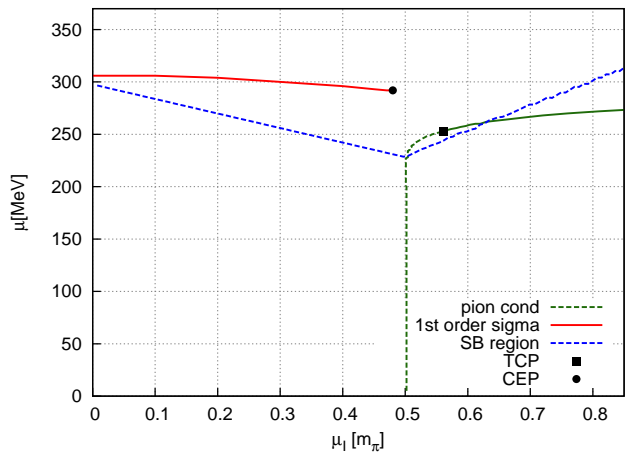}
\vspace{-.4cm}
  \caption{($\mu_I,\mu$) phase diagram at $T=0$ from an eMF calculation
    (parameter set A).  The SB line marks the boundary of the
    first/second Silver Blaze region.} 
  \label{fig:phase_diagram_t=0}
\vspace{-.2cm}
 \end{figure}

\paragraph{eMF calculation}
 \Fig{fig:phase_diagram_t=0} shows the phase diagram in the
 ($\mu_I,\mu$) plane from an eMF calculation. As discussed above,
 there is a chiral first order transition outside the pion
 condensation phase for large baryon chemical potential and small
 isospin chemical potential. With parameter set A, the 
 first order transition lies completely outside the first Silver Blaze region
 bounded by $\mu+ \mu_I < m_q$. The vertical line in the phase
 diagram separates the charged pion condensation phase from the normal chiral
 symmetry breaking phase. This phase boundary stays at the constant 
 $\mu_I = m_{\pi}/2$ until $\mu > m_q - m_{\pi}/2$ from where on it  
 bends to larger values of $\mu_I$ when further increasing $\mu$. 
For $\mu_I>m_\pi/2$ the general arguments from above require the phase
boundary of the pion condensation phase, as long as it is of second
order, to stay outside the second Silver Blaze region. Indeed, in
perfect agreement with these general arguments, the  
boundary of the pion condensation phase crosses the
Silver Blaze line only after it has become of first order, beyond a
tricritical point (TCP) as in Ref.~\cite{Andersen:2007qv},
which is here observed at $(\mu_I,\mu)=(0.56\,
m_\pi,253\, \text{MeV})$. We will see, however, that this
first order line gets washed out by the mesonic fluctuations. 

   \begin{figure}[t]
    \includegraphics[width=0.97\columnwidth]{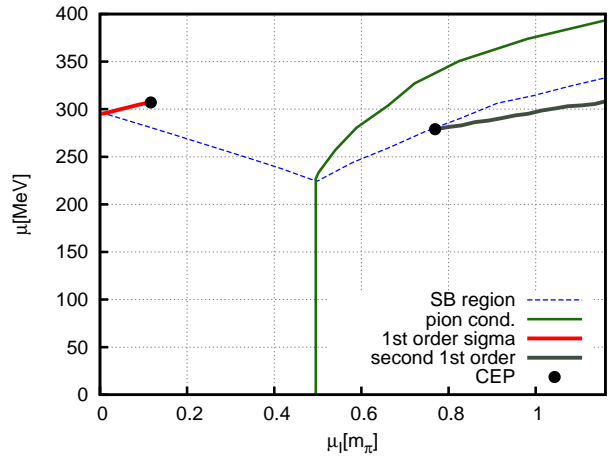}
\vspace{-.4cm}
    \caption{($\mu_I,\mu$) phase diagram at $T=0$ from the full
    FRG calculation (parameter set C). The SB line represents $\mu = m_{q}^-$.}
    \label{fig:phase_diagram_t=0_frg}
\vspace{-.2cm}
   \end{figure}

At the mean-field level the Silver-Blaze property is manifest in
the explicit expressions for the partition function. This has been
observed in \cite{Sasaki:2010jz} as a property of the zero temperature
partition function. It also holds for the explicit eMF expression for the 
effective potential \Eq{eq:effective action for MF}: one 
verifies that it remains constant throughout the first Silver Blaze
region with  $\mu + \mu_I  < m_{q}$ and $ \mu_I < m_{\pi} $.  
Moreover, with \Eq{eq:effective action for MF}, 
the condition for the second-order phase boundary of charged
pion condensation, 
\begin{equation}
 \begin{split}
  \frac{\partial U}{\partial d^2}(\mu_I^c)\Bigg|_{\rho^2=f_\pi^2,\, d^2 = 0} = 0,
  \label{eq:condition for 2nd order}
 \end{split}
\end{equation}
which defines the critical isospin chemical potential $\mu_I^c(\mu)$,
is in fact independent of $\mu$ for $\mu + \mu_I  \le g f_{\pi} \equiv
m_{q}$. This implies that the boundary of the pion condensation
phase stays constant at $\mu_{I}^c = m_{\pi}/2$ for
$\mu<m_{q} - m_{\pi}/2 $ as also seen in \Fig{fig:phase_diagram_t=0}.   
In fact, \Eq{eq:condition for 2nd order} coincides with 
 \begin{equation}
   \begin{split}
    \Gamma^{(0,2)}_{\pi}(p;\sigma)\Big|_{-p^2=(2\mu_I^c)^2,\, \sigma=f_\pi} = 0,
    \label{eq:condition for pole mass}
   \end{split}
 \end{equation}
from \Eq{eq:pion 2pt for MF} there. This establishes explicitly that
the pole mass agrees with the onset of pion condensation at $\mu_I^c=
m_\pi/2$  for all $\mu<m_{q} - m_{\pi}/2 $ in the eMF calculation. 

\paragraph{FRG calculation} \Fig{fig:phase_diagram_t=0_frg} shows the
corresponding phase diagram from the full FRG calculation. As in 
mean-field calculations there is a first order transition at
large $\mu$ and small $\mu_I$. However, the mesonic fluctuations tend
to weaken this first order transition and correspondingly the CEP is at
a smaller $\mu_I$ than in eMF. For small chemical potential $\mu$, the
second order transition separating charged pion condensation from the
normal $\chi$SB phase again occurs at $\mu_I = m_{\pi}/2$. And again,
the phase boundary stays constant until it hits the first Silver Blaze
boundary line at $\mu = m_q - m_{\pi}/2$, which defines the edge
between the first and the second Silver Blaze region.

   In contrast to mean-field calculations, however, the pion condensation
   transition with the mesonic fluctuations remains second order
   for the whole range of isospin chemical potentials we have
   investigated. Note also that this second order phase transition
   occurs outside the Silver Blaze region. Interestingly, one observes
   another first order transition inside this region for large isospin
   chemical potentials which was not present in the eMF
   calculation. This implies that the zero temperature partition function and
   correspondingly also the chiral and charged pion condensates  at
   fixed $\mu_I$ stay independent of $\mu$ only below this first order
   transition line which seems to end in a CEP right at the boundary
   of the second Silver Blaze region.

\subsection{Three dimensional ($\mu_I,\mu,T$) phase diagram}
\label{subsec:3dPD}

\begin{figure}[t]
    \includegraphics[width=0.97\columnwidth]{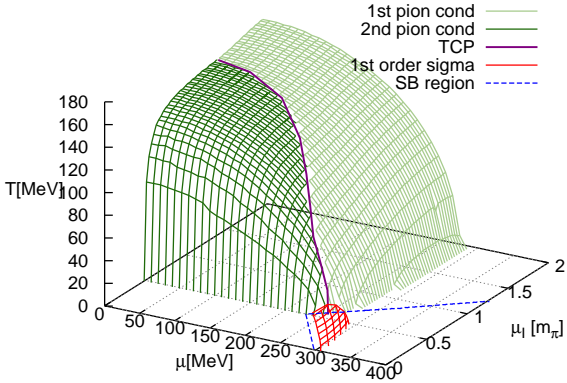}
\vspace{-.4cm}
    \caption{Three dimensional phase diagram from eMF (parameter set A).} 
    \label{fig:phase_diagram_3d}
   \end{figure}

\begin{figure}[t]
    \includegraphics[width=0.97\columnwidth]{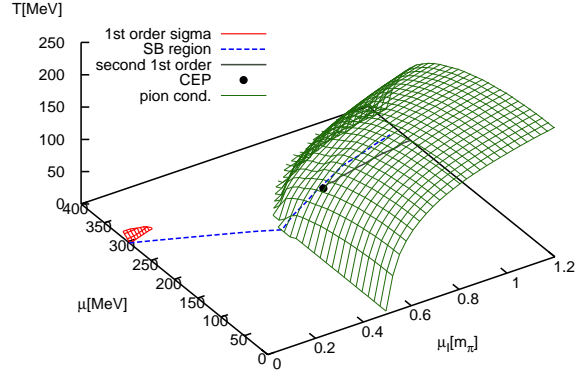}
\vspace{-.4cm}
    \caption{Three dimensional phase diagram from full FRG (parameter set C).}
    \label{fig:phase_diagram_3d_frg}
\vspace{-.4cm}
   \end{figure}

At finite temperature the phase boundaries in
Figs.~(\ref{fig:phase_diagram_t=0}) and 
(\ref{fig:phase_diagram_t=0_frg}) extend into surfaces in the three
dimensional parameter space with temperature $T$ over the ($\mu_I,\mu$) plane.  
These are shown for a representative eMF calculation in
\Fig{fig:phase_diagram_3d}, and a full FRG solution on the
two-dimensional grid in field space with fluctuations in the chiral as
well as the charged pion condensate in
\Fig{fig:phase_diagram_3d_frg}.

The green surfaces in these figures represent the boundary of the pion
condensation phase. The zero temperature TCP of
\Fig{fig:phase_diagram_t=0} in the eMF calculation 
extends into the purple line in \Fig{fig:phase_diagram_3d}, which
divides the surface of the pion condensation transition into a 
second order region (dark green, smaller $\mu_I$) and a first order
region (light green, larger $\mu_I$). At the other end of this purple
tricritical line, in the ($\mu_I,T$) plane at $\mu=0$, it becomes the
analogue of a TCP that was also predicted for two-color QCD at finite
baryon chemical potential  from next-to-leading order $\chi$PT
\cite{Splittorff:2002xn}. For a better comparison with the
corresponding QMD model mean-field results in 
\cite{Strodthoff:2011tz},
here we have removed the UV cutoff in the UV-finite thermal
contributions to the fermionic flow. When a common UV cutoff scale is
used for all fermionic fluctuations as in 
Eq.~(\ref{eq:effective action for MF}), on the other hand, this end
of the tricritical line bends towards larger $\mu_I$ and 
away from the $\mu=0$ plane. Including the full mesonic fluctuations
of chiral and charged pion condensates we observe no first order pion
condensation transition at all. The whole (light green)
first order surface towards the larger $\mu_I$ values is gone, see 
\Fig{fig:phase_diagram_3d_frg}, and with it the entire tricritical line. 
For $\mu = 0$ this is parallels the two-color case,
in which the fluctuations due to collective mesonic and baryonic excitations 
were found to remove the tricritical point and the corresponding
first order line \cite{Strodthoff:2011tz}. 

The chiral first order line of the $T=0$ eMF phase diagram for small
$\mu_I$, just outside the first Silver Blaze region in
\Fig{fig:phase_diagram_t=0}, is the baseline of the red surface in the
corresponding three-dimensional eMF phase diagram in
\Fig{fig:phase_diagram_3d}. Its edge 
represents the line of the critical endpoint in
\Fig{fig:phase_diagram_muf=0} which moves to lower temperature with
increasing $\mu_I$ ending up as that in \Fig{fig:phase_diagram_t=0}.  
As one can see from these figures, it thereby stays well clear of
the pion condensation phase.  The corresponding (red) first order
surface with fluctuations in \Fig{fig:phase_diagram_3d_frg} 
shows a similar behavior, albeit being much smaller than in the
eMF calculation as already observed at $T=0$. This first order
transition does not disappear but is considerably weakened by the
fluctuations as well. 
   
With fluctuations, there is another first order surface
inside the pion condensation phase. This is the extension  into the
three dimensional parameter space with temperature of the first
order line near the boundary of the second Silver Blaze region  in
\Fig{fig:phase_diagram_t=0_frg}. We have not investigated it in much
detail here and it is not included in \Fig{fig:phase_diagram_3d_frg},  
but it rises only to relatively small temperatures ending in
another line of critical endpoints similar to that of chiral first
order transition, but entirely contained inside the pion condensation
phase.

\begin{figure}[t]
     \includegraphics[width=0.97\columnwidth]{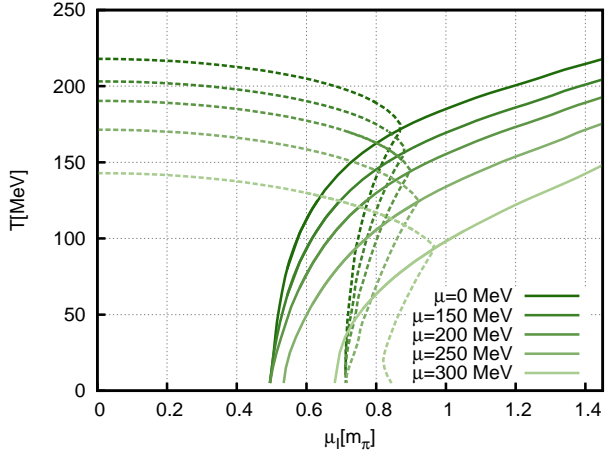}
\vspace{-.2cm}
     \caption{($\mu_I,T$) slices for different baryon chemical
       potentials of the three-dimensional FRG phase diagram in
       \Fig{fig:phase_diagram_3d_frg}. Also shown here are chiral
       crossover lines (dashed) as the half-value of the chiral condensate.} 
     \label{fig:phase_diagram_mu=0_frg}
    \end{figure}

Finally, in \Fig{fig:phase_diagram_mu=0_frg} we show various
($\mu_I,T$) planes for different values of the isosymmetric quark
chemical potential $\mu$ of the FRG phase diagram with fluctuations in
\Fig{fig:phase_diagram_3d_frg}.  The $\mu=0$ plane thereby corresponds
to the ($\mu,T$) phase diagram of the QMD model for two-color QCD with
diquark instead of charged pion condensation. This naturally reflects the 
precise map between the corresponding flow
equations. \Fig{fig:phase_diagram_mu=0_frg} also shows how the 
chiral condensate vanishes as the vacuum rotates 
in the pion condensation phase. For small baryon chemical potentials,
below the edge of the Silver Blaze regions at $\mu = m_q - m_\pi/2$,
small quantitative changes arise at finite $T$. At larger $\mu$ the
imbalance between the Fermi surfaces of up and anti-down quarks builds
up, and the pion condensation phase starts to shrink rapidly, likely
leading into an FFLO phase or alike in some possibly small region
where both  $\mu$ and $\mu_I$ are about equally large.

\vspace{-.3cm}

\section{Summary and Conclusions}
    \label{sec:conclusion}

\vspace{-.2cm}

We have investigated the phase diagram of the two-flavor QM
model for QCD with both isospin and baryon chemical potential using
the FRG. We showed explicitly how this effective model can be mapped
onto the QMD model for two-color QCD. As compared to earlier NJL model
studies, mesonic fluctuations are included with the FRG. We have
demonstrated their effects on the phase diagram by detailed comparisons
of purely fermionic flows in the eMF approximation with full FRG
results. As compared to previous FRG studies, collective mesonic
excitations are included in a way suitable to simultaneously describe
the competing fluctuations in both, the chiral and the charged pion
condensates, based on recent developments within the QMD model for
two-color QCD. As compared to the latter, a finite baryon chemical
potential here adds another dimension to the parameter space of the
phase diagram. Without baryon chemical potential, our results compare
well with recent lattice data indicating that the sigma might be
important for the meson mixing and the change in the low-lying 
excitation spectrum over the BEC-BCS crossover in the pion
condensation phase. At zero temperature, baryon chemical potentials
below generalized Silver Blaze bounds have no effect. Beyond those,
the increasing imbalance of the up and anti-down Fermi energies
reduces pion condensation as expected. Moreover, the onset of pion
condensation once more demonstrates how important it is to use proper
pole masses rather than screening masses for a realistic parameter
fixing in QM models. We have shown how to obtain good estimates for
those from the FRG in a way that is consistent with the FRG
calculation of the grand potential.

\vspace{-.3cm}

\section*{Acknowledgements}

\vspace{-.2cm}

     \noindent K.K.\ was supported by the Grant-in-Aid for JSPS
     Fellows (No. 22-3671) and by that for the Global COE Program "The
     Next Generation of Physics, Spun from Universality and
     Emergence." This work was supported by the Helmholtz
     International Center for FAIR within the LOEWE program of the
     State of Hesse, the Helmholtz Association Grant VH-NG-332, and
     the European Commission, FP7-PEOPLE-2009-RG No. 249203.  

\vspace{-.3cm}


\end{document}